\documentclass[twocolumn,prb,aps,footinbib,amsmath,amssymb,superscriptaddress]{revtex4-1}
\usepackage{graphicx}
\usepackage{dcolumn}
\usepackage{xcolor}
\usepackage{bm}
\usepackage{soul}
\usepackage{natbib,hyperref}
\usepackage{amsmath,amssymb,amsfonts,bbold}
\usepackage[normalem]{ulem}

\hypersetup{
	colorlinks=true,       
	linkcolor=blue,        
	citecolor=blue,        
	filecolor=magenta,     
	urlcolor=blue         
}

\begin{document}

\title{Two molecular devices for superconducting spintronics}

\author{Cristina Mier}
\affiliation{Centro de F{\'{i}}sica de Materiales
        CFM/MPC (CSIC-UPV/EHU),  20018 Donostia-San Sebasti\'an, Spain}
\affiliation{Donostia International Physics Center (DIPC),  20018 Donostia-San Sebasti\'an, Spain}
\author{{Alex} F\'etida}
\affiliation{Institut de Phsyique et Chime de Matériaux de Strasbourg, CNRS, Strasbourg, C\'edex, France} 
\author{Roberto Robles}
\affiliation{Centro de F{\'{i}}sica de Materiales
        CFM/MPC (CSIC-UPV/EHU),  20018 Donostia-San Sebasti\'an, Spain}
\author{Parmenio Boronat}
\affiliation{Centro de F{\'{i}}sica de Materiales
        CFM/MPC (CSIC-UPV/EHU),  20018 Donostia-San Sebasti\'an, Spain}
\affiliation{Institut de Ciència del Materials, ICMUV, 46980 Paterna, València, Spain}
\author{Divya Jyoti}
\affiliation{Centro de F{\'{i}}sica de Materiales
        CFM/MPC (CSIC-UPV/EHU),  20018 Donostia-San Sebasti\'an, Spain}
\affiliation{Donostia International Physics Center (DIPC),  20018 Donostia-San Sebasti\'an, Spain} 
\author{Nicol{\'a}s Lorente}
\affiliation{Centro de F{\'{i}}sica de Materiales
        CFM/MPC (CSIC-UPV/EHU),  20018 Donostia-San Sebasti\'an, Spain}
\affiliation{Donostia International Physics Center (DIPC),  20018 Donostia-San Sebasti\'an, Spain}    
\author{Laurent Limot}
\affiliation{Institut de Phsyique et Chime de Matériaux de Strasbourg, CNRS, Strasbourg, C\'edex, France} 
\author{Deung-Jang Choi}
\email{djchoi@dipc.org}
\affiliation{Centro de F{\'{i}}sica de Materiales
        CFM/MPC (CSIC-UPV/EHU),  20018 Donostia-San Sebasti\'an, Spain}
\affiliation{Donostia International Physics Center (DIPC),  20018 Donostia-San Sebasti\'an, Spain}
\affiliation{Ikerbasque, Basque Foundation for Science, 48013 Bilbao, Spain}

\begin{abstract}
    We create two molecular devices with superconducting junctions, using nickelocene molecules, single Fe atoms, and Pb electrodes at low temperature. We find contrasting behavior based on the coordination of the Fe atom: one device shows low-bias features in its differential conductance due to the superposition of multiple Andreev reflections (MAR) and Fe-induced in-gap states. The other reveals interference between MAR and in-gap states, showcasing the diversity achievable in atomically engineered devices with identical components.
\end{abstract}
\date{\today}

\maketitle

Superconductivity is a cornerstone in condensed matter physics and technology both for its conceptual importance as well as for its potential to drive new research~\cite{Tinkham}. Recent developments are showing two promising trends that integrate spintronics and superconductivity. The first direction focuses on spin-polarized current injection~\cite{Linder}, while the second explores non-reciprocal charge transport for rectification purposes~\cite{Ando, Nadeem}. And indeed the impact of these new trends can have important consequences in spin-based information technology~\cite{cai_2023}. Recently, it has been possible to probe the atomic limit of non-reciprocal charge transport and rectification, resulting in the first atomic-scale superconducting diode~\cite{Katharina_2023}. This last example underscores the captivating strategy of combining atomic spins and superconducting junctions  for probing the atomic limits of radically new quantum devices. 

Among atomic quantum devices, molecule-based devices offer reproducibility and functionality. In conjunction with superconductivity, interesting effects appear. A notable example is the enhancement of spin excitation lifetimes, a key parameter in quantum technologies. Recent experiments have revealed that molecular magnets exhibit long lifetimes on a superconductor, attributed to the gapped quasiparticle spectra characteristic of superconductors~\cite{BenjaminH}. The magnetic properties of these molecular junctions can vary considerably, influenced by factors like molecular magnetic anisotropy and the nature of the electrode coupling~\cite{Hatter, Laetitia}. Furthermore, the high density of states at the superconducting gap edge amplifies inelastic conductivity signatures in transport measurements. This enhancement becomes instrumental in distinguishing molecular spin processes~\cite{Liljeroth, Mier_JPCL, Drost_2023}.

\begin{figure}[b!]
\begin{center}
\includegraphics[width=.75\linewidth]{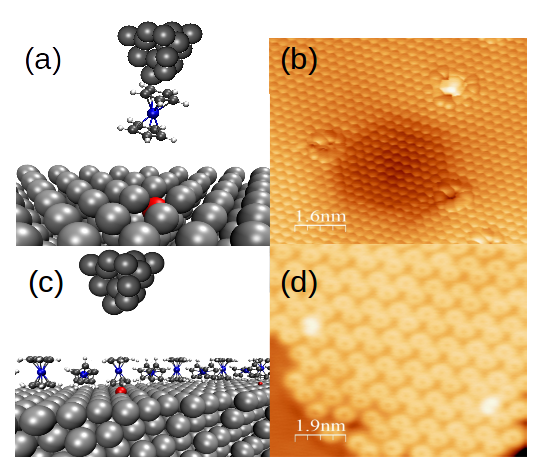}
\end{center}
\caption{Two distinct molecular devices. Shown in (a), atomic scheme of a  Fe atom (red circle) underneath the first atomic layer of Pb (111) addressed by a Nc-molecule-terminated Pb-coated tip. The Nc molecule contains two C$_5$H$_5$ rings and a Ni atom (blue).  (b) Constant-current image of three of these subsurface Fe atoms close to an Ar-induced defect~\cite{Ar_Pb111} ($\mbox{I}=200$~pA, $\mbox{V}=10$~mV). The second device (c) is a Pb-terminated tip positioned on a Nc monolayer where the Fe impurity (red) is located at the molecule/Pb (111) interface. (d) The constant-current image ($\mbox{I}=30$~pA, $\mbox{V}=-30$~mV) shows Nc dimers~\protect{\cite{Bachellier,Mier_JPCL}}, where two Fe atoms are located underneath two Nc molecules (brighter spots).}
\label{Fig1}
\end{figure}

In this study, we demonstrate the effective integration of molecules into superconducting molecular devices. Our approach involved the construction of two distinct devices, each comprising a single magnetic molecule, nickelocene (Nc, formed by two cyclo-pentadyenil rings, C$_5$H$_5$, and an inner nickel atom), and an iron (Fe) atom, positioned between two lead (Pb) superconducting electrodes in a scanning tunneling microscope (STM) setup. These junctions exhibit weak-link characteristics~\cite{WeakLinks, Kroeger_2017} and achieve stability through the inclusion of the molecule~\cite{SI}. 

The weak link in these devices is characterized by distinct tunneling phenomena: Cooper-pair tunneling manifesting as the Josephson effect at zero bias and multiple Andreev reflection (MAR) processes at finite bias. Such phenomena are not unique to our system but have been observed in other STM superconducting configurations as well~\cite{Naaman1, Naaman2, Suderow, Ternes, Randeria, Jack_2015, Laetitia, Karan_2022}. Notably, in our experiments, identical components were employed to construct different atomic-scale devices, each distinguished by unique atomic arrangements. This resulted in devices with markedly different transport properties.

\begin{figure*} 
\hspace*{-0.3cm}
\centering
\includegraphics[width=0.9\linewidth]{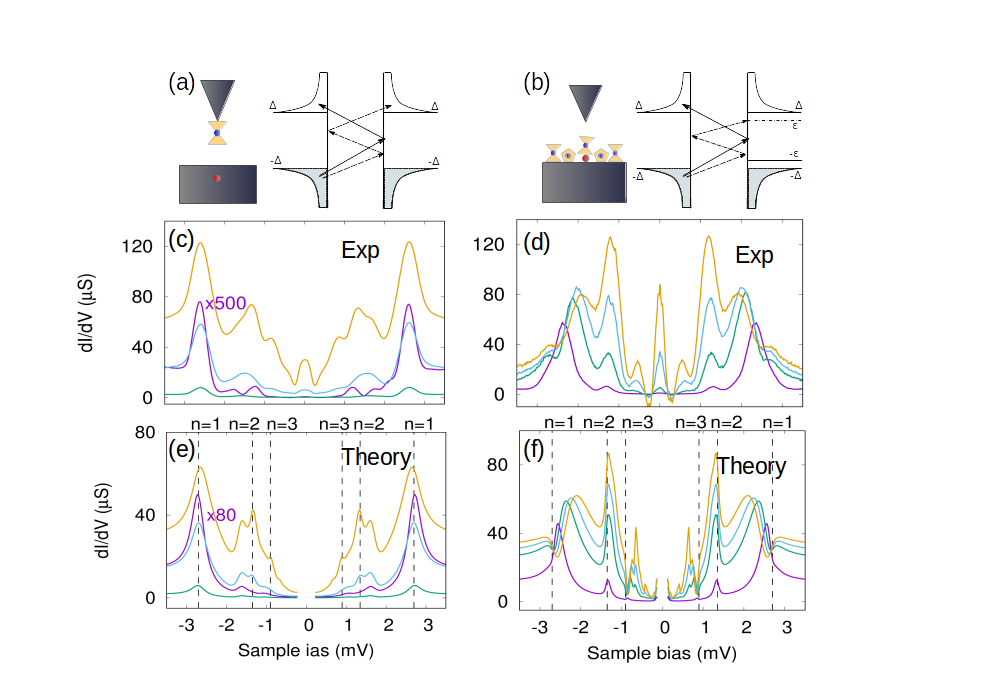}
\caption{
Differential conductance of the two different devices. The schemes (a) and (b) show the two molecular junctions together with the MAR. In (b) $\varepsilon$ refers to the binding energy of the in-gap state. (c) Differential conductance of the subsurface Fe atom where a molecule tip is approached leading to set-point conductances  $G = 0.04, 2.0, 20, 60$ $\mu$S (purple, green, cyan, yellow). The purple dI/dV signal ($G = 0.04$ $\mu$S) is multiplied by the marked factor for clarity. (d) Differential conductance obtained on the Fe below the molecular layer for set-point conductances $G = 10, 20, 30, 40$ $\mu$S (purple, green, cyan, yellow). As in (c), purple corresponds to the lower transparency and yellow to the higher transparency of the junction. (e) and (f) are the calculated conductances (color code follows the same trend as in the experiment), where the dashed lines correspond to pure MAR processes given by $(\Delta+\Delta_{tip})/n\approx2\Delta/n=2.7/n$~meV. In (e), we can observe in-gap states induced by the sub-surface Fe atom, while in (f) we see split-off peaks for odd-numbered MAR showing the composition of in-gap and MAR in the detected peaks.  Due to the truncation in Floquet terms and the absence of fluctuating phases, the very-low bias region is not computed.} 
\label{Fig2}
\end{figure*}

To conduct our measurements, we used an STM at a base temperature of 2.5 K under ultra-high-vacuum conditions. We evaporated the Nc molecules onto the Pb~(111) surface resulting in self-assembled molecular layers~\cite{SI}. We used a Pb-coated STM tip,
leading to a gap of  $\sim2\Delta$ in the differential conductance as a function of bias ($\Delta+\Delta_{tip}\approx 2\Delta$). 
The procedure was completed with the in-situ Fe evaporation, reaching a coverage of less than 1\% of individual Fe atoms on the  Pb surface.

Utilizing the atomic-manipulation capabilities of our STM, we constructed the first device depicted in Fig.~\ref{Fig1} (a). This was achieved by first picking up one of the Nc molecules from the surface using the Pb-coated tip and then approaching it to an Fe atom. The incorporation of the molecule at the tip's apex~\cite{Mier_JPCL} allowed for atomic resolution imaging of the Pb (111) surface, Fig.~\ref{Fig1} (b). The close match between the experimental and computed STM images using density functional theory (DFT)~\cite{SI} confirms that the Fe atoms are  subsurface, in agreement with previous observations for other transition-metal atoms on Pb surfaces~\cite{Katharina_2023,Ruby_2015,Choi_2017}. 

The other device, Fig.~\ref{Fig1} (c), is created by bringing the STM tip close to the Nc ontop of a Fe atom, seen as bright spots in Fig.~\ref{Fig1} (d). The new device consists of two superconducting electrodes (tip and surface) and a Nc-Fe junction where the Nc molecule is coupled to the tip  and the Fe is coupled to the surface. 
Furthermore, both the constant-current image of Fig.~\ref{Fig1} (d) and our DFT calculations show that the lower cyclo-pentadienyl ring increases its overlap with the Fe atom creating a strong bond that prevents the Fe atom from slipping beneath the Pb~(111) surface~\cite{SI}. Consequently, the tunneling electrons from the Pb-coated tip are able to interact with the superconducting surface, which includes Fe-induced in-gap states~\cite{Randeria} within the molecular junction.

In Figure ~\ref{Fig2}, panels (a) and (b) provide schematic illustrations of the two devices. Panel (a) depicts a subsurface Fe atom, denoted by a red dot, beneath the superconducting substrate. This atom is probed by a {Nc-terminated} STM tip, represented as a pair of cyclo-pentadienyl  rings in yellow, connected by a Ni atom, shown as a blue dot. On the right of this illustration, an energy versus distance diagram presents a one-electron view of the electron-transfer processes at the tunnel junction, in analogy with established representations of electron tunneling in superconducting systems~\cite{Tinkham,cuevas_full_2003}. In this configuration, the magnetic impurity, being remote from the surface, does not disrupt the MAR processes. Conversely, panel (b) illustrates the scenario where an Fe atom is positioned between the molecule and the substrate. Here, the presence of the magnetic impurity on the surface modifies the in-gap states, which in turn, interacts with MAR processes as depicted by the more intricate scheme. These unconventional MAR processes can be pictured as excitations of the in-gap states by electrons or holes that are Andreev reflected from the superconductor~\cite{Paaske_2011,Randeria,Laetitia,Karan_2022,Cuevas_2020}.

The experimental implications of these two distinct configurations are captured in panels (c) and (d). Each panel exhibits four differential conductance ($dI/dV$) curves as a function of the sample bias for various STM set points. These set points range from a tunneling regime {(illustrated in purple)}, predominantly displaying the quasiparticle peak at $\sim 2\Delta$, along with minor in-gap features, to higher differential conductance (shown in green, cyan, and yellow as the conductance increases) where in-gap structures become more pronounced. 

{The low-bias electron-transport behaviors of the two devices are starkly different.  Indeed, at large junction resistance (purple in Fig.~\ref{Fig2} (c) and (d)), both curves show a clear quasiparticle peak at $2\Delta$ and some smaller features near $\Delta$.  Increasing the junction conductance by bringing the tip closer modifies these features differently for the two devices: for (c), the \(2\Delta\) peaks are mostly stable, getting broader; for (d), however, the peak splits giving a shoulder at \(2\Delta\) and a peak shifting to lower bias. In case (c), the intensity of these peaks invariably increases with conductance, whereas in case (d), the intensity of the split peak initially increases but subsequently decreases, exhibiting an unexpected non-monotonic behavior. The next peak at lower bias exhibits a consistent growth in magnitude with increased conductance in both cases, a trend that is also observed for smaller bias. Ultimately, a pronounced Josephson-like feature is observed, which can be accounted for in terms of the thermal fluctuations of the superconducting phases~\cite{Naaman1,Naaman2,Steiner_2023,SI}.}


To gain a deeper understanding of the observed phenomena, we have formulated two models comprising two BCS superconductors linked via a molecular bridge, here the Nc molecule. The molecular bridge is accounted for by an electron transmission, denoted as \(\alpha\), and the Fe atom is simplified to a Kondo Hamiltonian. In this notation, \(J\) represents the Kondo exchange coupling constant, while \(S\) denotes the spin of the Fe atom, which is treated as a classical variable. The dynamics of the system is solved using a Floquet expansion of the two-time non-equilibrium Green's functions within Nambu space~\cite{SI,cuevas_hamiltonian_1996,Paaske_2011,Carlos}. The difference between the two models is how the Fe atom relates to the Nc transmission~\cite{SI}. For the first device, the Fe atom affects the transmission in the bulk, while for the second device, the Fe affects the transmission in the junction.

The theoretical results, depicted in Fig.~\ref{Fig2} panels (e) and (f), show good agreement with the experimental data if the process of moving the STM tip closer not only increases the electron transmission coefficient \(\alpha\) but also the exchange interaction \(J\). By varying these two parameters within our model, we are able to understand the peak evolution of the experimental graphs, Figs~\ref{Fig2} (c) and (d).

Figure~\ref{Fig2} (e) corresponds to a model of a tunneling junction characterized by $\alpha$ that is further coupled with a second transmission to a bulk impurity. The bulk impurity yields in-gap states via $JS$. At low $\alpha$, the full electron transmission can be seen as the contribution of two coherent electronic steps: firstly, the transmission to the substrate with MAR depending on \(\alpha\); secondly, the transmission through the substrate modified by the in-gap states introduced by the bulk impurity~\cite{SI}. 

As a consequence, Fig.~\ref{Fig2} (e) shows the superposition of the clean-substrate MAR (dashed lines for the first three MAR orders) together with the in-gap states. In agreement with Fig.~\ref{Fig2} (c), the  peak at \(2\Delta\) is stationary because it corresponds to the MAR \(n=1\) peak, independent of the subsurface Fe atom. In contrast, the succeeding peak at a lower voltage originates from an in-gap state solely produced by the Fe impurity. Experimentally, the  peak at \(V \approx 1.15\) mV for lower conductance (purple line) corresponds to a thermal replica of this Fe-induced in-gap state, easily confirmed by our theoretical results. As conductance increases, this feature is overshadowed by the more rapidly intensifying \(n=3\) MAR peak. This analysis shows that the present Nc molecular bridge at 2.5~K allows us to reveal three orders of MAR.

Figure~\ref{Fig2} (f) illustrates a model in which MAR processes incorporate in-gap states, as detailed in Refs.~\cite{Randeria,Cuevas_2020,SI}. Notably, even at lower tunneling rates, the peak near $2\Delta$ exhibits a discernible shift and distortion. This peak splits into two distinct components as the tunneling rate increases: one remains anchored at $2\Delta$, while the other evolves in response to an increase in $JS$. To reproduce our experimental observations, we increase $JS$ concomitant with the enhancement in electron transmission $\alpha$. Specifically, the second MAR peak (n=2) shows a pronounced dependency on $\alpha$, overcoming the in-gap state split from $2\Delta$. 
The $n=2$ peak reveals a composite nature, comprising in-gap state contributions. Similarly, the third MAR peak (n=3) mirrors the behavior of the first (n=1), forming a sharp shoulder. This leads us to identify an odd-even effect in the MAR peaks: odd-numbered peaks (n=1, 3, ...) exhibit in-gap contributions  causing additional broadening and distortion of the low-bias peaks. In contrast, even-numbered peaks remain stationary, 
with more intensity and broadening as the junction's transparency increases. The root of this odd-even effect lies in the alignment of the MAR's final state with an in-gap state for odd-ordered reflections. 

\begin{figure}[t!]
\hspace*{-0.5cm}
\includegraphics[width=1.1\linewidth]{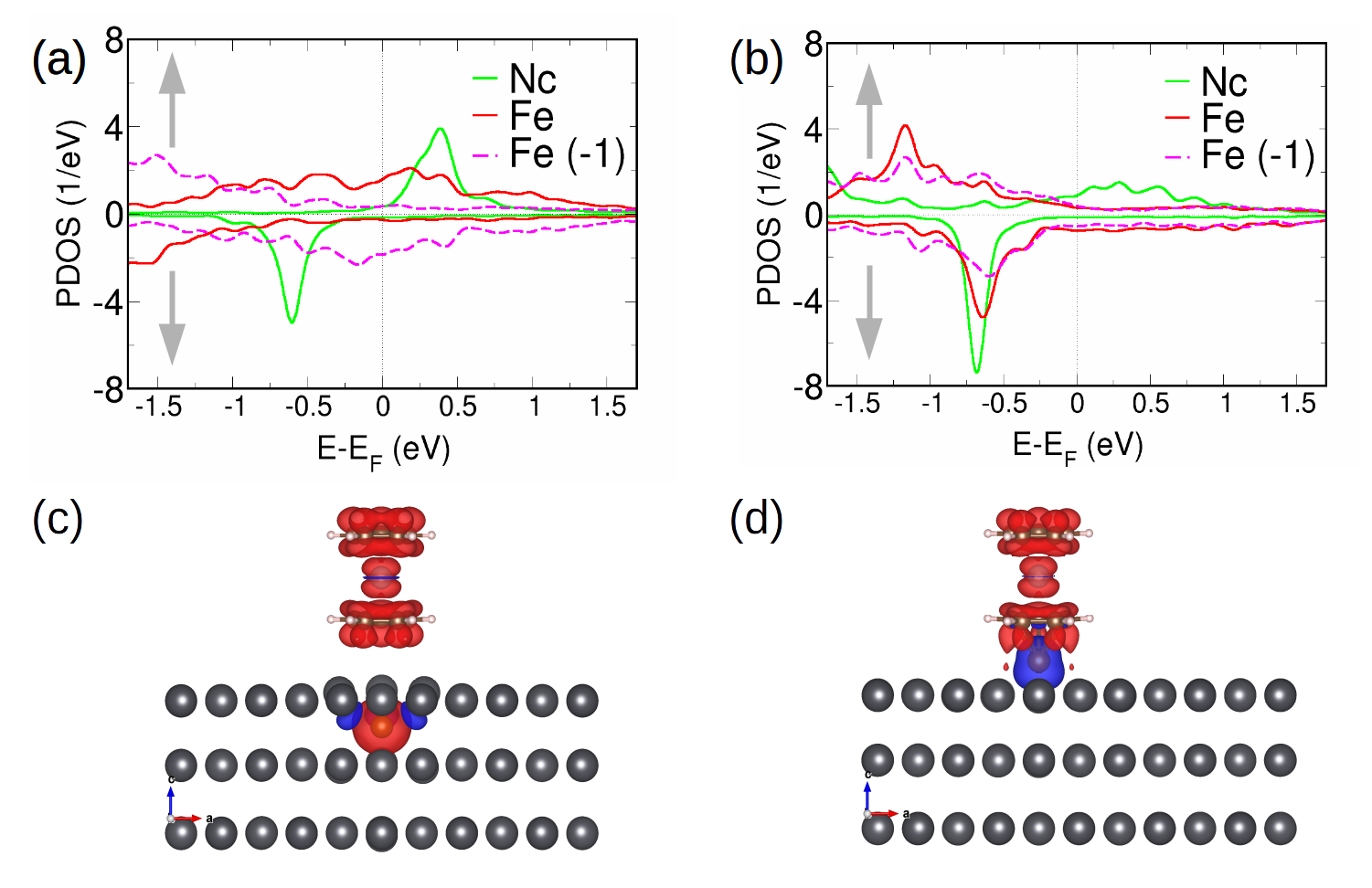}
\caption{ Density-functional theory results.
The projected  density of states (PDOS) on the molecular states (green, Nc) and on the (a) subsurface or (b) interface Fe atom (red, Fe) corresponding to the atomic schemes in (c) and (d), respectively.
The PDOS for the first device (a) clearly shows the alignment of majority and minority spins between Fe and Nc, in agreement with the spin density in (c), where the red isosurface shows the up-spin density and blue the down-spin one.
When the top cyclopentadienyl ring of the molecule is rigidly displaced 1 \AA~towards the surface (equivalent to reducing the junction gap in 1~\AA ), the Fe atom changes its spin (magenta dashed lines, Fe (-1)).
In the second device (b) we see that the majority-spin PDOS have opposite signs corresponding to the antiferromagnetic configuration shown in (d).
}
\label{Fig4}
\end{figure}

Density Functional Theory (DFT) calculations enable us to elucidate the differences between the two devices~\cite{SI}. Figure~\ref{Fig4} (a) and (b) show the projected density of states (PDOS) on the molecular states (represented in green for Nc) and on the Fe atom (in red for Fe), corresponding to the molecule adsorbed vertically as per the atomic schemes of the two devices in (c) and (d) respectively.

The PDOS for the first device (a) distinctly reveals the alignment of majority and minority spins between Fe and Nc, resulting in a ferromagnetic configuration. This is exemplified in the spin density in (c), where red and blue isosurfaces represent up-spin and down-spin densities, respectively. Here, the Fe and Nc molecules interact indirectly through the first Pb layer, as reflected in their mutual magnetic ordering. 
Compressing the molecule by 1 \AA, a notable spin inversion occurs in the Fe atom (indicated by magenta dashed lines for Fe (-1)).
These results indicate that, despite the close proximity of Fe and the molecule in the first device, significant structural compression is required for direct interaction.

Figures~\ref{Fig4} (b) and (d) pertain to the interface-Fe device. 
Here, Fe and the molecule maintain an antiferromagnetic configuration, even under 1~\AA{} compression, showing that 
Fe and Nc are in direct interaction. Consequently, increasing their mutual interaction via molecular compression does not alter their mutual magnetic ordering but leads to significant broadening of the spin-up frontier orbitals of Nc. 

In conclusion, the integration of Nc molecules, Fe adatoms, and Pb  electrodes yields distinctly different electronic transport behaviors, critically dependent on the positioning of the Fe adatom relative to the Nc molecule. The computed diffusion barrier for Fe adsorbed on the surface is about 50 meV, leading to its subsurface localization, as corroborated by experimental imagery~\cite{SI}. This results in the Fe atom being encapsulated by the Pb layer, which attenuates the direct coupling to the Nc tip. Consequently, in this configuration, the Nc molecule functions as a non-magnetic intermediary in the superconducting electrodes, while the Fe atoms contribute in-gap states that are extrinsic to the MAR processes.

In stark contrast, when the Nc molecule is adsorbed onto the surface prior to Fe atom addition, the Fe atom tends to nestle between the Nc molecule and the surface, forming a robust bond. This setup gives rise to a significantly diminished total magnetic moment, with the Nc and Fe elements aligning in an antiferromagnetic fashion. This pronounced interaction between the Fe and Nc alters the weak link considerably, with the in-gap states induced by the Fe atom directly influencing the MAR processes, thereby manifesting a distinctly different signature from that observed in the prior configuration.

These findings underscore the feasibility of engineering molecular systems that alter the properties of superconducting devices. Notably, molecules embodying multiple magnetic centers can substantially expand the design parameters essential for the advancement of superconducting spintronic applications.

We thank projects  PID2021-127917NB-I00 by MCIN/AEI/10.13039/501100011033, QUAN-000021-01 by Gipuzkoa Provincial Council, IT-1527-22 by Basque Government, 202260I187 by CSIC, ESiM project 101046364 by EU, Marie Skłodowska-Curie grant 847471 by EU, the International Center for Frontier Research in Chemistry (Strasbourg), and computational resources by Finisterrae II (CESGA). Views and opinions expressed are however those of the author(s) only and do not necessarily reflect those of the EU. Neither the EU nor the granting authority can be held responsible for them. 

\newpage

\nocite{Ormaza_2017,Harada_1996,Johnson_noise, Nyquist,Mier2_2021,moire,vonOppen_2020,Tunnel2,Ester2,Salkola_1997,Shiba1,Moca_2008,Alfredo_1995,Keldysh1,Keldysh2,Keldysh3,Keldysh4,Dynes,Kresse1996b,Kresse1999,PBE,D3BJa,D3BJb,tersoff_theory_1985,bocquet_theory_2009,lorente_stmpw_2019}
\bibliography{references}
\end{document}